\documentclass[prl,superscriptaddress,showpacs,longbibliography,reprint]{revtex4-2} 

\usepackage[colorlinks=true,linkcolor=blue,urlcolor=blue,citecolor=blue,pdfusetitle]{hyperref}
\usepackage{color}
\usepackage[utf8]{inputenc}
\usepackage[usenames,dvipsnames]{xcolor}
\usepackage{amsthm}
\usepackage{amsmath}
\usepackage{amssymb}
\usepackage{graphicx}
\usepackage{bm}
\usepackage[all]{xy}
\usepackage{lipsum}
\usepackage{braket}
\usepackage{dsfont}
\usepackage{array}
\usepackage{makecell}
\usepackage{xr}
\usepackage{float}

\newtheorem{theorem}{Theorem}
\newtheorem{lemma}{Lemma}

\begin{document}
\raggedbottom

\title{Applicability of mean-field theory for time-dependent open quantum systems with infinite-range interactions}

\newcommand{\tubingen}{Institut f\"{u}r Theoretische Physik,  Universit\"{a}t T\"{u}bingen, Auf der Morgenstelle 14, 72076 T\"{u}bingen, Germany}

\author{Federico Carollo}
\affiliation{\tubingen}
\author{Igor Lesanovsky}
\affiliation{\tubingen}
\affiliation{School of Physics and Astronomy and Centre for the Mathematics and Theoretical Physics of Quantum Non-Equilibrium Systems, The University of Nottingham, Nottingham, NG7 2RD, United Kingdom}

\date{\today}

\begin{abstract}
Understanding quantum many-body systems with long-range or infinite-range interactions is of relevance across a broad set of physical disciplines, including quantum optics, nuclear magnetic resonance and nuclear physics. From a theoretical viewpoint, these systems are appealing since they can be efficiently studied with numerics, and in the thermodynamic limit are expected to be governed by mean-field equations of motion. Over the past years the capabilities to experimentally create long-range interacting systems have dramatically improved permitting their control in space and time. This allows to induce and explore a plethora of nonequilibrium dynamical phases, including time-crystals and even chaotic regimes. However, establishing the emergence of these phases from numerical simulations turns out to be surprisingly challenging. This difficulty led to the assertion that mean-field theory may  not be applicable to time-dependent infinite-range  interacting systems. Here, we rigorously prove that mean-field theory in fact exactly captures their dynamics, in the thermodynamic limit. We further provide bounds for finite-size effects and their dependence on the evolution time.
\end{abstract}

\maketitle
\noindent \textbf{Introduction.---} 
A multitude of currently investigated many-body quantum systems features effective infinite-range interactions 
 \cite{solano2017,muniz2020,mivehvar2021,ferioli2023,defenu2023}. 
This interaction type dramatically influences both the equilibrium \cite{kastner2010,kastner2010b,russomanno2021,botzung2021,defenu2024} and the nonequilibrium \cite{kastner2011,eisert2013,defenu2018,defenu2021,bachelard2013,bhakuni2021,defenu2023} properties of quantum and classical \cite{dauxois2002,bouchet2010} many-body systems.
From a theoretical perspective, infinite-range interacting systems are often described through collective (spin) models, which is appealing since they can be efficiently simulated numerically \cite{chase2008,baragiola2010,kirton2017,Shammah2018}. Moreover, in the thermodynamic limit, mean-field theory is usually applicable (see Refs.~\cite{fowler2023,carollo2023} for a counterexample). Here, interactions lead to effective nonlinearities in the equations of motion \cite{hepp1973,hepp1973b,spohn1980,alicki1983,benedikter2016,benatti2016,benatti2018,carollo2021,fiorelli2023}. 

\begin{figure}[t!]
    \centering
    \includegraphics[width=\columnwidth]{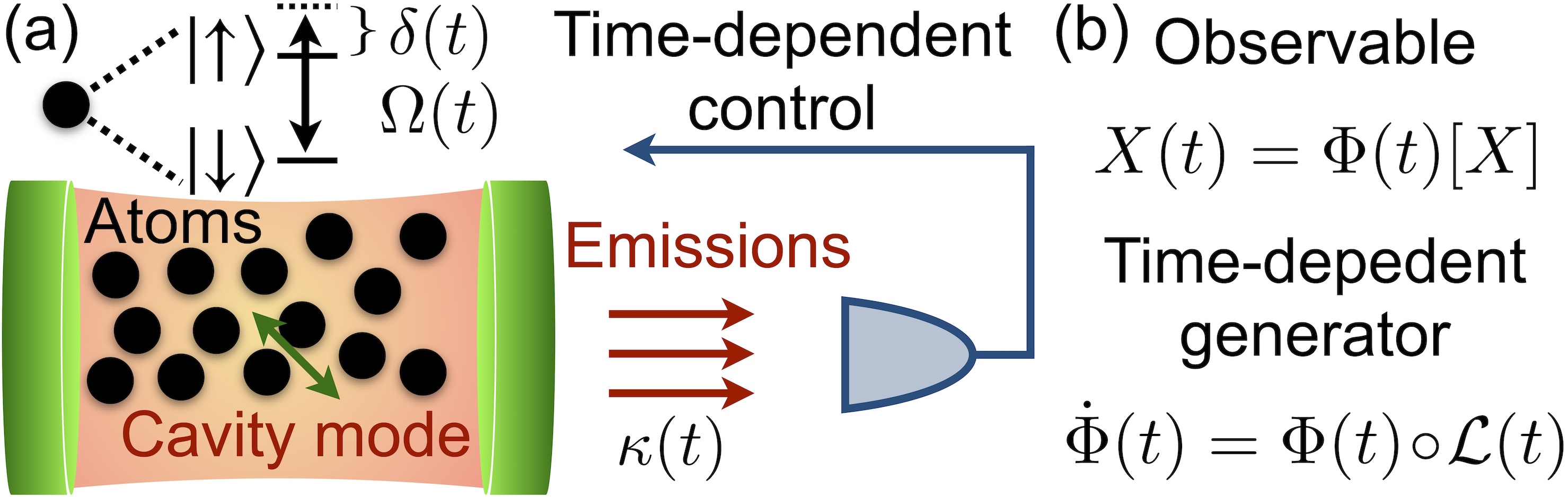}
    \caption{{\bf Time-dependent driving and control in an atom-cavity setup.} (a) Atoms (described by spin-$1/2$ particles) are coupled to a single cavity mode. The collective atom-cavity interaction gives rise to an effective infinite-range interacting system. Experiments allow to dynamically change the Rabi frequency, $\Omega(t)$, and the detuning, $\delta(t)$, of a laser driving the atoms. Even photo-emission rates may be varied in time and emissions from the cavity may be used to apply control operations conditional on the emission signal. (b) Driven-dissipative protocols like the one sketched in panel (a) generically induce non-Markovian effects in the evolution of a quantum observable, $X(t)=\Phi(t)[X]$, implemented by the propagator $\Phi(t)$. In these instances, the propagator features a generator $\mathcal{L}(t)$ [cf.~Eq.~\eqref{generator}], which is explicitly time-dependent.}
    \label{fig0}
\end{figure}

A paradigmatic platform featuring infinite-range interactions is an atom-cavity system, which is sketched in Fig.~\ref{fig0}(a). Here, an ensemble of atoms inside the cavity couples collectively with a single optical cavity mode, which mediates long-range interactions \cite{agarwal1997,gopalakrishnan2011,norcia2018,palacino2021}. Current experiments allow for the dynamical control of interaction strengths and dissipation rates. Moreover, the information gathered from monitoring emissions (see, e.g., Ref.~\cite{wiseman2009}) permits conditional time-dependent driving, which depends on the emission signal itself, as in the case of continuous feedback \cite{wiseman1994,lammers2016,ivanov2020,kroeger2020,ivanov2021}.

The combination of these techniques thus allows to realize open quantum infinite-range interacting many-body systems, whose dynamics is governed by time-dependent generators, see Fig.~\ref{fig0}(b). Such {\it time-local} generators capture a large class of problems, since they can also represent generic non-Markovian effects, including those expressed through memory kernels \cite{chruscinski2010}. Within a mean-field description, systems governed by time-dependent generators show interesting nonequilibrium phases \cite{ivanov2020,ivanov2021,zhihao2024}. However, establishing their emergence can be rather controversial, since numerical results  may look incompatible with the mean-field prediction. This discrepancy has even led to the assertion that mean-field theory fails to capture the dynamics of infinite-range interacting systems undergoing strong driving \cite{zhihao2024}. 

In this paper, we resolve this issue by rigorously proving that mean-field theory is exact, in the thermodynamic limit, for infinite-range interacting systems featuring time-dependent dynamical generators [cf.~Fig.~\ref{fig0}(b)].  Our analytical derivation  further shows how to interpret finite-size numerical results, which are affected by the interplay between the  size of the system and the length of the evolution time. To benchmark our findings, we reconsider the dissipative Floquet Ising model of Ref.~\cite{zhihao2024} and investigate the dynamics of collective observables through extensive simulations. Despite the presence of large finite-size effects, we show that numerical results are indeed compatible with mean-field theory. 
Finally, we analyze the spectrum of the microscopic evolution map,  which allows us to link the  spectral gap with the emergence of different  mean-field phases. Our work rigorously establishes mean-field theory as an efficient and faithful approach for the study of explicitly time-dependent and non-Markovian dynamics in contemporary experimental setups. 
\\

\noindent \textbf{Infinite-range time-dependent dynamics.---} We consider quantum systems consisting of $N$ $d$-level particles, with $d<\infty$.  The single-particle algebra is spanned by a  Hermitean basis $\left\{v_\alpha \right\}_{\alpha=1}^{d^2}$, such that $v_\alpha^\dagger=v_\alpha$ and ${\rm Tr}(v_\alpha v_\beta)=\delta_{\alpha\beta}$. Commutators between basis elements can be written in terms of the structure factors $\varepsilon_{\mu\nu}^\eta$, as  $[v_\mu,v_\nu]=i\sum_{\eta}\varepsilon_{\mu\nu}^\eta v_\eta$.  Since we are interested in models with infinite-range interactions, we introduce the collective operators $V_\alpha=\sum_{k=1}^N v_\alpha^{(k)}$, where $v_\alpha^{(k)}$ indicates the operator $v_\alpha$ of the $k$th particle. 
These systems are subject to an open quantum dynamics (see Fig. \ref{fig0}) described by the propagator $\Phi(t)$, which implements the dynamics of any operator $X$ as $X(t)=\Phi(t)[X]$. The propagator is defined as the solution of the equation $\dot{\Phi}(t)=\Phi(t)\circ\mathcal{L}(t)$, with $\circ$ denoting composition of maps and with time-dependent generator $\mathcal{L}(t)$ having  the form \cite{lindblad1976,breuer2002}
\begin{equation}
\begin{split}
    \mathcal{L}(t)[\cdot]&=i[H(t),\cdot]+\sum_{\alpha,\beta}\frac{c_{\alpha \beta}(t)}{2N} \left(\left[V_\alpha,\cdot\right]V_\beta+V_\alpha\left[\cdot,V_\beta\right]\right)\, ,\\
    &H(t)=\sum_{\alpha} \omega_\alpha(t) V_\alpha+\sum_{\alpha,\beta} \frac{h_{\alpha\beta}(t)}{N}V_\alpha V_\beta \, . 
    \end{split}
    \label{generator}
\end{equation}
The operator $H(t)=H^\dagger(t)$ represents the Hamiltonian of the system. The frequencies $\omega_\alpha(t)$ specify the noninteracting term of the Hamiltonian, while the matrix $h(t)$ encodes the structure of the two-body infinite-range interactions. We consider $h(t)$  to be real. This is because imaginary coefficients of $h(t)$ would lead to commutators between collective operators, resulting in irrelevant (in the thermodynamic limit) noninteracting terms rescaled by $1/N$.  The matrix $c(t)$ in Eq.~\eqref{generator}, which we decompose as $c(t)=a(t)+ib(t)$, must also be Hermitean. If it is also positive [$c(t)\ge0$] for all times, the dynamics is Markovian, in the sense of being {\it divisible} in terms of completely positive maps \cite{breuer2009,rivas2010,rivas2014,chruscinski2014,devega2017}. 
In general, $c(t)$ can feature negative eigenvalues at certain times and $\mathcal{L}(t)$ can thus encode non-Markovian effects. Even in these cases, the propagator $\Phi(t)$ must be a completely positive map since it  represents a physical dynamics. 
For the sake of concreteness, we assume all entries of $\omega(t)$, $h(t)$ and $c(t)$ to be analytic functions on the positive real line $t\ge0$ (see Supplemental Material \cite{SM} \vphantom{\cite{perko13,teschl2012,coddington1956}} for details). The rescaling $1/N$ in front of the ``quadratic" terms in  Eq.~\eqref{generator} ensures a well-defined thermodynamic limit ($N\to\infty$). 
\\

\noindent \textbf{Emergent mean-field theory.---} An important class of observables for many-body (open) quantum systems is that of macroscopic observables \cite{benatti2018,lanford1969}. The latter are defined as $m_\alpha^N=V_\alpha/N$ and encode macroscopic properties of the system. These operators provide relevant order parameters able to discriminate between different many-body phases.  
To derive the dynamics of macroscopic operators, one computes the action of the generator on them, $\mathcal{L}(t)[m_\alpha^N]$. The resulting operators consist of second-order polynomials of macroscopic operators \cite{SM}. Applying the generator on these gives rise to a hierarchy of Heisenberg equations, which are not closed in the large-$N$ limit. To make analytical progress, one typically resorts to  so-called mean-field equations. The latter  are obtained under the assumption that the operators $m_\alpha^N$ become, in the thermodynamic limit, multiples of the identity. The equations read \cite{SM}
\begin{equation}
    \dot{m}_\alpha(t)= \sum_{\beta=1}^{d^2} \zeta_\beta(t) m_\beta(t)+\sum_{\beta,\beta'=1}^{d^2}\xi_{\beta\beta'}(t) m_\beta(t)m_{\beta'}(t) \, ,
    \label{mf-eqs}
\end{equation}
with coefficients $\zeta_{\beta}(t)=-\sum_{\mu}\omega_\mu(t)\varepsilon_{\mu \alpha}^\beta$ and $\xi_{\beta\beta'}(t)=-\sum_{\mu}\left[b_{\mu \beta'}(t)+2 h_{\mu\beta'}(t)\right] \varepsilon_{\mu\alpha}^{\beta}$. In Eq.~\eqref{mf-eqs},  $m_\alpha(t)$ is expected to approximate the behavior of $m_\alpha^N$ in the large-$N$ limit. 
In what follows, we demonstrate that such an approximation is in fact exact for systems described by the time-dependent generator in Eq.~\eqref{generator}. More precisely, we prove that
\begin{equation}
    \lim_{N\to\infty}\langle \Phi(t)[m_\alpha^N]\rangle=m_\alpha(t)\, ,
    \label{limit}
\end{equation} 
with $m_\alpha(t)$ being the solution of Eq.~\eqref{mf-eqs}. 

To this end, we introduce a suitable  ``error function",
\begin{equation}
    \mathcal{E}_N(t):=\sum_{\alpha=1}^{d^2} \left\langle \Phi(t) \left[\left(m_\alpha^N-m_\alpha(t)\right)^2\right]\right\rangle\, ,
    \label{error}
\end{equation}
which measures how much the macroscopic operators $m_\alpha^N$ deviate, at time $t$, from $m_\alpha(t)$. The rationale is that showing $\lim_{N\to\infty}\mathcal{E}_N(t)=0$ directly implies Eq.~\eqref{limit} and, thus, the validity of the mean-field theory in Eq.~\eqref{mf-eqs}. 
As shown by the theorem below, the value assumed by the error function $\mathcal{E}_N(t)$, at any time $t$, can be controlled by its initial value $\mathcal{E}_N(0)$. Thus, if the initial state is such that $\mathcal{E}_N(0)\to0$, the mean-field approximation is exact.

\begin{theorem}
\label{th1}
Considering the generator in Eq.~\eqref{generator} and the assumption on its coefficients, one has that 
\begin{equation}
\label{GL}
\mathcal{E}_N(t)\le e^{C_1  t}\mathcal{E}_N(0)+\frac{C_2}{C_1 N}\left(e^{tC_1}-1\right)\, ,
\end{equation}
where $C_1,C_2$ are $N$-independent constants. Moreover, if the initial   state is such that $\lim_{N\to\infty}\mathcal{E}_N(0)=0$, then $\lim_{N\to\infty}\mathcal{E}_N(t)=0$, for any $t>0$. 
\end{theorem}
The proof of the theorem is given in the Supplemental Material \cite{SM}. It relies on showing that the function $\mathcal{E}_N(t)$ is  differentiable and it exploits the approach put forward in Ref.~\cite{carollo2021} to derive the bound in  Eq.~\eqref{GL}. 
Typical initial states of theoretical and experimental interest feature  short-range correlations, e.g., uncorrelated product states. For such states,  $\mathcal{E}_N(0)\sim 1/N$, so that 
\begin{equation}
    \mathcal{E}_N(t)< \frac{e^{C_1 t}}{N}\left(1+\frac{C_2}{C_1}\right)\, .
    \label{Boound}
\end{equation}
The above inequality leads to two related statements: $i)$ For any finite system, the mean-field equations may become inaccurate at sufficiently large times; $ii)$ For any finite time $t$,  the mean-field equations become more and more accurate upon increasing the system size $N$. The larger is the time $t$ the larger is the system size required to reach a desired level of accuracy. These observations are crucial for recognizing signatures of emergent mean-field dynamics in finite-size analysis. 
\\

\begin{figure*}[t!]
    \centering
    \includegraphics[width=\textwidth]{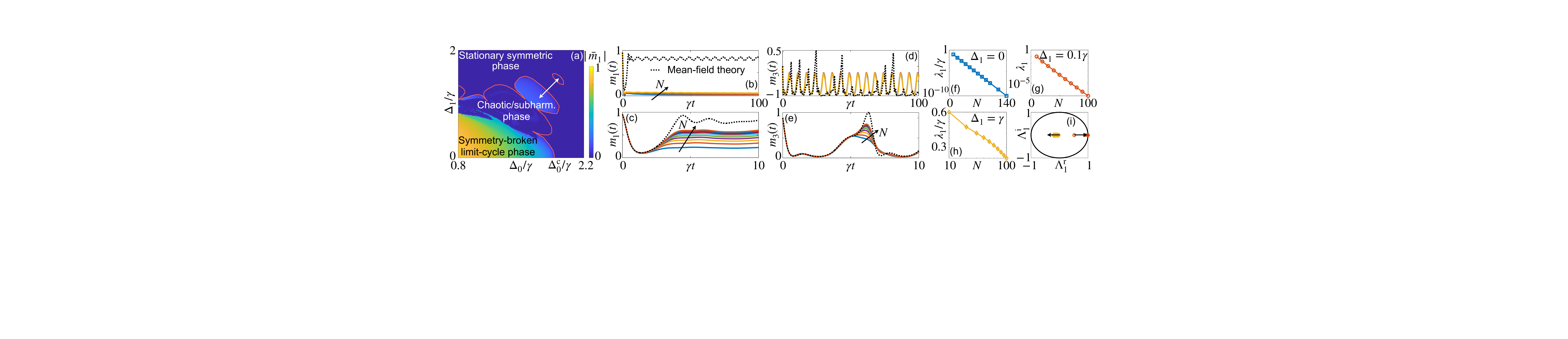}
    \caption{{\bf Dissipative Floquet Ising model.} (a)  Order parameter $|\bar{m}_1|$,  where $\bar{m}_1=[\int_0^{n\tau} {\rm d}t \, m_1(t)]/(n\tau)$, as a function of $\Delta_0/\gamma$ and $\Delta_1/\gamma$ with $n=300$. 
    (b-c) Comparison between the mean-field magnetization of $m_1(t)$ and finite-$N$ exact results for  $\Delta_1/\gamma=0.1$ and $\Delta_0/\gamma=1$. Panel (b) shows results for $N=10,20,30$. Panel (c) shows instead a more extensive comparison for $N=200, 400,\dots, 1800$ in a smaller time window. (d-e) Same as in (b-c) for $\Delta_1/\gamma=1$. (f) Semi-log plot of the gap $\lambda_1$ as a function of the system size for the time-independent case $\Delta_1=0$. (g) Same as in (f) for $\Delta_1/\gamma=0.1$. (h) Log-log plot of the gap as a function of the system size for $\Delta_1=\gamma$. (i) Plot of the eigenvalue $\Lambda_1$ in the complex plane for $\Delta_1=0.1\gamma$ (positive real part) and for $\Delta_1=\gamma$ (negative real part). In all panels, the remaining dynamical parameters are $g=\chi=\gamma$ and the initial state is characterized by $\theta=0.5\pi$,  $\phi=0.1\pi$.  }
    \label{fig1}
\end{figure*}

\noindent \textbf{Dissipative Ising model.---} To explore how mean-field behavior emerges from finite systems, we consider an infinite-range dissipative quantum Ising model. The collective operators for the model are the operators  $V_\alpha=\sum_{k=1}^N\sigma_\alpha^{(k)}$ ($\alpha=1,2,3$), where $\sigma_\alpha$ are Pauli matrices. (Note that the matrices $\sigma_\alpha$ are not normalized, since ${\rm Tr}(\sigma_\alpha\sigma_\beta)=2\delta_{\alpha\beta}$, but this is not essential for the proof.)
The dynamical generator has the form in Eq.~\eqref{generator}, with  time-periodic Hamiltonian
\begin{equation}
    H(t)=-g\frac{V_1^2}{N}+\left[\Delta_0 + \Delta_1 \sin (\chi t)\right] V_3\, ,
    \label{Ising_Hamiltonian}
\end{equation}
and time-independent dissipation described by the matrix $c_{11}=c_{22}=2\gamma$ and $c_{12}=-c_{21}=i2\gamma$. Such a dissipative contribution accounts for collective decay [through the jump operator $V_-=(V_1-iV_2)/(2\sqrt{N})$] at rate $2\gamma$ \cite{SM}.
In Eq.~\eqref{Ising_Hamiltonian}, the parameter $g$ encodes the Ising coupling strength, while the external field (in the $z$-direction) consists of a constant term $\Delta_0$ and a time-dependent driving, with amplitude $\Delta_1$ and frequency $\chi$. As initial state, we consider fully polarized product states, i.e., with spins all pointing in the same direction. The initial expectation is given by $\langle \cdot\rangle=\bra{\psi(0)}\cdot \ket{\psi(0)}$, with $\ket{\psi(0)}=\bigotimes_{k=1}^N\ket{\theta,\phi}$, where $\theta,\phi$ are Euler angles. The infinite-range dissipative Ising model falls in the class of systems in Eq.~\eqref{generator} and the initial state is such that  $\lim_{N\to\infty}\mathcal{E}_N(0)=0$. We can thus investigate the emergent behavior of the model via a mean-field theory. \\

\noindent \textbf{Nonequilibrium  phase diagram.---} In the absence of the external field ($\Delta_0=\Delta_1=0$), the dissipative Ising model introduced above shows a nonequilibrium transition from a stationary phase to a boundary time-crystal \cite{iemini2018,buca2019,carollo2022}, featuring self-sustained oscillations out of a static driving. However, such a phase does not manifest for $\Delta_0>0$. From now on, we focus on finite $\Delta_0$ values [see Fig.~\ref{fig1}(a)] and on the largest angular momentum sector.

Both the system Hamiltonian $H(t)$ and the dissipative term feature a $Z_2$-symmetry, $V_1\to-V_1$ and $V_2\to-V_2$, implemented by the parity operator $\prod_{k=1}^N\sigma_3^{(k)}$.  In the absence of periodic driving ($\Delta_1=0$) and for sufficiently large external field $\Delta_0>\Delta_0^{\rm c}\approx 1.87$ [cf.~Fig.~\ref{fig1}(a)], the system features a unique stationary state, which possesses the same symmetry of the dynamical generator. On the other hand, for sufficiently weak external fields $\Delta_0<\Delta_0^{\rm c}$, the system spontaneously breaks the symmetry of the generator and enters a ``stationary" bistable phase. The stable solutions present a nonzero magnetization $m_1$, with opposite sign. For a given choice of Euler angles for the initial state $\ket{\psi(0)}$, the system approaches either one or the other symmetry-broken state. 

When switching on the field $\Delta_1>0$, the system Hamiltonian becomes periodic in time. Due to the presence of dissipation, one expects the system to approach, at long times, a unique limit cycle, with same period of the driving, $\tau=2\pi/\chi$. This is what is observed generically for any finite system. However, in the thermodynamic limit, we observe an interesting phase diagram. For small driving amplitudes $\Delta_1$, moving away from the symmetric stationary phase, the macroscopic magnetizations remain stationary even in the presence of the periodic driving. On the other hand, turning on the periodic driving starting from the stationary bistable regime $\Delta_0<\Delta_0^{\rm c}$, leads to the emergence of two stable symmetry-broken limit-cycle solutions. The latter essentially originate from small oscillations around the  stable stationary solutions. Which limit cycle is approached depends on the initial state. 

For increasingly large parameter $\Delta_1$ in the region $\Delta_0<\Delta_0^{\rm c}$, an interesting phenomenology emerges. The system starts showing subharmonic responses to the  time-dependent driving, such as, for instance,  a limit cycle with period $2\tau$ \cite{zhihao2024}. For even larger amplitudes $\Delta_1$, there can emerge even lower harmonic responses until the system enters a fully chaotic regime \cite{zhihao2024}. If $\Delta_1$ is further increased above a certain threshold [$\Delta_1\approx 1.5$ in Fig.~\ref{fig1}(a)], the periodic driving brings the system into the symmetric stationary phase. \\

\noindent \textbf{Finite-size analysis.---} The prediction from mean-field theory is strictly valid only in the thermodynamic limit $N\to\infty$, see, e.g., Eq.~\eqref{Boound}. In realistic experiments, quantum systems can be very large but they are anyway always finite. In what follows, we perform extensive numerical investigations and observe and discuss signatures of emergent mean-field behavior in finite systems.

We start by looking at the dynamics of macroscopic magnetization operators in the symmetry-broken limit-cycle phase. In Fig.~\ref{fig1}(b), we show a comparison between mean-field theory and simulations for small systems. The mean-field prediction approaches a limit-cycle with positive $m_1(t)$. For relatively small systems, we observe instead a convergence toward $m_1=0$, which is necessary since 
for any finite system the stationary state is symmetric. We note that this approach to stationarity becomes slower and slower the larger the system size. At first sight, these results seemingly suggest that mean-field theory fails to capture the behavior of the system. However, one has to remember in which sense mean-field theory is valid. Its validity requires looking at a given time  and taking the limit $N\to\infty$ [cf.~Eq.~\eqref{Boound}]. This scaling is presented in Fig.~\ref{fig1}(c), which shows indeed how finite-$N$ simulations approach the mean-field prediction. 

Next, we consider a point in the phase diagram associated with chaotic dynamics. As shown in Fig.~\ref{fig1}(d), while the mean-field prediction shows an aperiodic pattern, finite-$N$ results approach a periodic solution,  with period $\tau$. Also this plot may seem to indicate a failure of the mean-field theory (as stated in Ref.~\cite{zhihao2024}). However, an approach to a periodic solution for large times is necessary for any finite systems, since the latter always feature a unique asymptotic limit-cycle state. To see that mean-field theory is exact also in this case, we consider, in Fig.~\ref{fig1}(e), larger system sizes within an initial time window [see again Eq.~\eqref{Boound}]. The agreement for increasing system sizes is evident.  \\

\noindent \textbf{Floquet spectrum and gap.---} We now connect the emergence of {\it exotic} mean-field solutions, with aperiodic behavior or with period different from  the driving period $\tau$, to properties of the dynamical generator. To this end, we consider the spectrum of the  Floquet map in the Schr\"odinger picture (see details in Ref.~\cite{SM}), $\Phi^*(\tau)$, which  evolves the system state  over a single period $\tau$. If such a map can be diagonalized, we can write it as  
\begin{equation}
    \Phi^*(\tau)[\rho]=\sum_m  \Lambda_m {\rm Tr}\left(\ell_m \rho\right)r_m\,, 
\end{equation}
with $r_m,\ell_m$ being the right and the left eigenmatrices of the  Floquet map and $\Lambda_m$ the associated eigenvalues [the spectrum of $\Phi(\tau)$ coincides with that of $\Phi^*(\tau)$]. Due to physical constraints, the map $\Phi^*(\tau)$ has an eigenvalue $\Lambda_0=1$, necessary for trace-preservation, while the remaining ones must be such that $|\Lambda_m|\le 1$. If $\Lambda_0$ is the sole eigenvalue with modulus  one, the system state converges, in the long-time limit, to a unique limit cycle, $\rho_{\rm LC}(t)$, $t\in[0,\tau)$. The whole limit cycle can be obtained by noticing that $\rho_{\rm LC}(0)=\lim_{n\to\infty}\Phi^{*\, n}(\tau)[\rho]=r_0$ and $\rho_{\rm LC}(t)=\Phi^*(t)[\rho_{\rm LC}(0)]$, $t\in[0,\tau)$. 

Nontrivial dynamical behavior, such as that in the phase diagram of Fig.~\ref{fig1}(a), emerges when additional eigenvalues approach a unit modulus, $|\Lambda_m|\to1$, in the thermodynamic limit. For instance, when the second dominant eigenvalue $\Lambda_1\to1$, the system still approaches a limit cycle with period $\tau$. However, the latter is degenerate as it depends on the initial state $\rho$, since 
\begin{equation}
\rho_{\rm LC}(0)\to r_0+{\rm Tr}[\ell_1 \rho(0)]r_1\, .
\label{spectrum_bistable}
\end{equation}
On the other hand, whenever $\Lambda_1\to -1$, the system enters a discrete time-crystalline phase (DTC) (see, e.g., Refs.~\cite{taheri2022,taheri2022b,vu2023}), with a degenerate limit cycle showing a period $\tau_{\rm DTC}=2\tau$, since
\begin{equation}
    \rho_{\rm DTC}(n\tau)\to r_0+ (-1)^n {\rm Tr}[\ell_1 \rho(0)] r_1\, .
    \label{dtc}
\end{equation}
The factor $(-1)^n$ enforces that the system returns to its current state only after two periods. Responses with larger multiples of the  driving period, or even aperiodic responses, require two or more additional  eigenvalues approaching unit modulus and with imaginary components. 

We now study the spectrum of the Floquet map in different scenarios, focusing in particular on the gap $\lambda_1=\log |\Lambda_1|/\tau$, which tends to $0$ when degenerate asymptotic solutions emerge.
In Fig.~\ref{fig1}(f), we show the gap for the  time-independent case $\Delta_1=0$, in the bistable regime. The gap closes exponentially with the system size, denoting the emergence of two stable solutions. In Fig.~\ref{fig1}(g), we consider the parameter regime of Fig.~\ref{fig1}(b-c) featuring a bistable limit-cycle phase. Also in this case, the gap closes exponentially and $\Lambda_1\to1$ [cf.~Fig.~\ref{fig1}(i) and Eq.~\eqref{spectrum_bistable}]. Finally, in Fig.~\ref{fig1}(h) we consider the case of Fig.~\ref{fig1}(d-e) showing the emergence of chaotic behavior. In this case, the gap closes with a power-law behavior $\lambda_1\propto N^{-1}$, with  $\Lambda_1\to -1$ [cf.~Fig.~\ref{fig1}(i)]. The fact that  mean-field theory shows aperiodic behavior implies that for even larger systems (beyond the sizes we can access) other eigenvalues $\Lambda_m$ must approach $|\Lambda_m|\to 1$. \\

\noindent \textbf{Discussion.---} We have presented a rigorous proof of the validity of  mean-field theory for infinite-range open quantum spin systems, described by a generic time-dependent dynamical  generator. This includes the situation of time-periodic, or Floquet, open quantum dynamics. Our proof can be extended to account for other types of dynamical effects, such as local dissipative processes \cite{roberts2023,fiorelli2023}, which leave the system permutation invariant. Our approach may also be exploited to prove the validity of mean-field theory in genuine stroboscopic dynamics, realized via piecewise continuous generators $\mathcal{L}(t)$. 

To benchmark our theory, we have performed extensive numerical simulations of a dissipative Floquet Ising model, which was also studied in Ref.~\cite{zhihao2024}. There it was suggested that mean-field theory fails to capture the dynamical behavior at strong driving ($\Delta_1=\gamma$). We were unable to confirm this and also did not find striking differences between treating the weak and strong driving regimes. Finite-size effects can, however, be sizable. Also the analysis of the spectrum of the Floquet propagator did not display any peculiarities that would hint towards a breakdown of mean-field theory in the thermodynamic limit. 
\\

\textbf{Acknowledgments.---} We acknowledge funding from the Deutsche Forschungsgemeinschaft (DFG, German Research Foundation) under Project No. 435696605 and through the Research Unit FOR 5413/1, Grant No. 465199066, and through the Research Unit FOR 5522/1, Grant No. 499180199. This project has also received funding from the European Union’s Horizon Europe research and innovation program under Grant Agreement No. 101046968 (BRISQ). F.C.~is indebted to the Baden-W\"urttemberg Stiftung for the financial support of this research project by the Eliteprogramme for Postdocs.

\bibliography{references}

\newpage
\setcounter{equation}{0}
\setcounter{figure}{0}
\setcounter{table}{0}
\makeatletter
\renewcommand{\theequation}{S\arabic{equation}}
\renewcommand{\thefigure}{S\arabic{figure}}
\makeatletter

\onecolumngrid
\newpage

\setcounter{page}{1}
\begin{center}
{\Large SUPPLEMENTAL MATERIAL}
\end{center}
\begin{center}
\vspace{0.8cm}
{\Large Applicability of mean-field theory for time-dependent open quantum systems with infinite-range interactions}
\end{center}
\begin{center}
Federico Carollo$^{1}$ and Igor Lesanovsky$^{1,2}$
\end{center}
\begin{center}
$^1${\em Institut f\"ur Theoretische Physik, Universit\"at T\"ubingen,}\\
{\em Auf der Morgenstelle 14, 72076 T\"ubingen, Germany}\\
$^2${\em School of Physics and Astronomy and Centre for the Mathematics}\\
{\em and Theoretical Physics of Quantum Non-Equilibrium Systems,}\\
{\em  The University of Nottingham, Nottingham, NG7 2RD, United Kingdom}\\

\end{center}

\section*{I. Main theorem}
In this section, we provide a step-by-step proof of the theorem stated in the main text. Before that, we start defining some important quantities that will be exploited later on. \\

We start by decomposing the generator in Eq.~\eqref{generator} of the main text into four different parts. The first two  account for the Hamiltonian contribution and are given by
$$
\mathcal{H}^{\rm Loc}(t)[X]=i\sum_{\mu}\omega_\mu(t) V_\mu \, , \qquad  \mathcal{H}^{\rm Int}(t)[X]=i\sum_{\mu\nu}\frac{h_{\mu\nu}(t)}{N}\left[V_\mu V_\nu,X\right]\, .
$$
The other two terms are instead related to dissipation and are defined as  
$$
\mathcal{A}(t)[X]=\sum_{\mu\nu}\frac{a_{\mu\nu}(t)}{2N}\left[\left[V_\mu,X\right],V_\nu\right]\, , \qquad \mathcal{B}(t)[X]=i\sum_{\mu\nu}\frac{b_{\mu\nu}(t)}{2N}\left\{\left[V_\mu,X\right],V_\nu\right\}\, ,
$$
with $a(t)$ and $b(t)$ being the real and the imaginary part of the Kossakowski matrix $c(t)$. Through these maps the generator can be written as  $\mathcal{L}(t)=\mathcal{H}^{\rm Loc}(t)+\mathcal{H}^{\rm Int}(t)+\mathcal{A}(t)+\mathcal{B}(t)$. \\

We then consider the action of these different terms on the macroscopic observables $m_\alpha^N=V_\alpha/N$ considered in the main text. For the Hamiltonian contributions, this gives 
\begin{equation}
\mathcal{H}^{\rm Loc}(t)[m_\alpha^N]=-\sum_{\mu,\eta}\omega_\mu(t) \varepsilon_{\mu\alpha}^\eta m_\eta^N\, ,\qquad \mathcal{H}^{\rm Int}(t)[m_\alpha^N]=-\sum_{\mu, \nu,\eta}h_{\mu\nu}(t)\left(\varepsilon_{\nu\alpha}^\eta m_\mu^N m_\eta^N+\varepsilon_{\mu\alpha}^\eta m_\eta^N m_\nu^N\right)\, ,
\label{rel1}
\end{equation}
where we made use of the structure constants $\varepsilon_{\mu\nu}^\eta$, such that $[V_\mu,V_\nu]=i\varepsilon_{\mu\nu}^\eta V_\eta$. Moreover, for the dissipative part we obtain 
\begin{equation}
\mathcal{A}(t)[m_\alpha^N]=-\sum_{\mu,\nu,\eta,\xi}\frac{a_{\mu\nu}(t)}{2N}\varepsilon_{\mu\alpha}^\eta\varepsilon_{\eta\nu}^\xi m_\xi^N\, ,\qquad \mathcal{B}(t)[m_\alpha^N]=-\sum_{\mu,\nu,\eta}\frac{b_{\mu\nu}(t)}{2}\varepsilon_{\mu\alpha}^\eta\left(m_\eta^N m_\nu^N+m_\nu^N m_\eta^N\right)\, .
\label{rel2}
\end{equation}
It is important to note that the map $\mathcal{A}(t)$ generates terms proportional to a single macroscopic observable further suppressed by a factor $1/N$. As we shall see, this contribution becomes irrelevant in the thermodynamic limit. The relations above allow us to directly write the mean-field equations reported in Eq.~\eqref{mf-eqs} of the main text. We indeed notice that $\dot{\Phi}(t)[m_\alpha^N]=\Phi(t)\circ\mathcal{L}(t)[m_\alpha^N]$ and assuming convergence of the macroscopic variables to multiples of the identity, we find 
\begin{equation}
\dot{m}_\alpha(t)=-\left[\sum_{\mu,\eta}\omega_\mu(t)\varepsilon_{\mu\alpha}^\eta m_\eta(t) +\sum_{\mu,\nu,\eta}h_{\mu\nu}(t)\left(\varepsilon_{\nu\alpha}^\eta m_\mu(t) m_\eta(t)+\varepsilon_{\mu\alpha}^\eta m_\eta(t) m_\nu(t)\right)+\sum_{\mu,\nu,\eta}b_{\mu\nu}(t)\varepsilon_{\mu\alpha}^\eta m_\eta(t) m_\nu(t)\right]\, ,
    \label{mf_SM}
\end{equation}
which can be rewritten as they appear in Eq.~\eqref{mf-eqs} of the main text. \\

We can now proceed with the actual proof of the theorem, recalling that the assumptions are: 
\begin{itemize}
    \item C1) The functions  $\omega_\mu(z),h_{\mu\nu}(z),c_{\mu\nu}(z)$ are analytic for $z\in D$, where $D$ is a simply connect domain in the complex plane $\mathbb{C}$. The domain contains,  in its interior, the non-negative part of the real line. 
    \item C2) The initial state of the system is such that  $\lim_{N\to\infty}\mathcal{E}_N(0)=0$. 
\end{itemize}

{\it Proof:} The idea of the proof is to find a suitable bound to the growth of the error function $\mathcal{E}_N(t)$ [cf.~Eq.~\eqref{error}] with time. To this end, it is convenient to look at the derivative of such a quantity. In Lemma 2, we show that the map $\Phi(t)$, solving the differential equation $\Phi(t)\circ\mathcal{L}(t)$, exists and is analytic in $D$ due to our assumption C1. In Lemma 1, we show that the solution to the mean-field equations exists and is a differentiable function over the whole (positive) real line. As a consequence of these two Lemmata (proven in the next Section), we conclude that the function $\mathcal{E}_N(t)$ is differentiable, for $t>0$. We thus have
\begin{equation}
\dot{\mathcal{E}}_N(t)=\sum_{\alpha=1}^{d^2} \left\{\left\langle \Phi(t)\circ\mathcal{L}(t)\left[\left(m_\alpha^N-m_\alpha(t)\right)^2\right]\right\rangle-2\dot{m}_\alpha(t)\left\langle \Phi(t)\left[m_\alpha^N-m_\alpha(t)\right]\right\rangle\right\}\, .
\label{S1}
\end{equation}
Note that, $m_\alpha(t)$ is a scalar quantity and can thus be freely moved inside and outside of the expectation value $\langle \cdot\rangle$. Moreover,  we have $\mathcal{L}(t)[m_\alpha(t)]=0$.
For our generator, we further have the following equality
$$
\mathcal{L}(t)[XY]=\mathcal{L}(t)[X]Y+X\mathcal{L}(t)[Y]+\sum_{\mu,\nu=1}^{d^2}\frac{c_{\mu\nu}(t)}{N}\left[V_\mu,X\right]\left[Y,V_\nu\right].
$$
Exploiting this relation in Eq.~\eqref{S1} allows us to write 
\begin{equation}
\begin{split}
\dot{\mathcal{E}}_N(t)&=\sum_{\alpha=1}^{d^2}\left\{\left\langle \Phi(t)\left[\Big(\mathcal{L}(t)\left[m_\alpha^N\right]-\dot{m}_\alpha(t)\Big)\Big(m_\alpha^N-m_\alpha(t)\Big)\right]\right\rangle+
\left\langle \Phi(t)\left[\Big(m_\alpha^N-m_\alpha(t)\Big)\Big(\mathcal{L}(t)\left[m_\alpha^N\right]-\dot{m}_\alpha(t)\Big)\right]\right\rangle\right\}\\
&+\sum_{\alpha=1}^{d^2}\sum_{\mu,\nu=1}^{d^2}\left\langle \Phi(t)\left[\frac{c_{\mu\nu}(t)}{N}\left[V_\mu,m_\alpha^N\right]\left[m_\alpha^N,V_\nu\right]\right]\right\rangle\, .
\end{split}
    \label{partial}
\end{equation}
The second term in the first line of the above equation is the complex conjugate of the first one, so that we only need to control one of the two. The third term (second line of the above equation), which we call here $T_3$ is instead bounded by 
\begin{equation}
\left|T_3\right|=\left|\sum_{\alpha=1}^{d^2}\sum_{\mu,\nu=1}^{d^2}\left\langle \Phi(t)\left[\frac{c_{\mu\nu}(t)}{N}\left[V_\mu,m_\alpha^N\right]\left[m_\alpha^N,V_\nu\right]\right]\right\rangle\right|\le \frac{4d^6 c_{\rm max}}{N}\, .
\label{bound_3}
\end{equation}
Here, we have assumed that the coefficients of the dynamical generator are all bounded on the non-negative real line by the constant $c_{\rm max}=\sup_{ t\in[0,\infty),\forall\mu,\nu}\left\{|c_{\mu\nu}(t)|,|h_{\mu\nu}(t)|,|\omega_\mu(t)|\right\}$ (see last paragraph of this Section for a discussion about those cases in which these coefficients are not bounded when $t\to\infty$). We further used that $\|v_\alpha\|\le 1$ and that $\left\|\left[V_\mu,m_\alpha^N\right]\right\|\le 2$.

We then consider the first sum in Eq.~\eqref{partial}. We focus on the term inside the first round brackets. Using the mean-field equations and the relations in Eqs.~\eqref{rel1}-\eqref{rel2}, we can write 
\begin{equation}
\begin{split}
\mathcal{L}(t)\left[m_\alpha^N\right]-\dot{m}_\alpha(t)&=-\sum_{\mu,\eta=1}^{d^2}\omega_{\mu}(t)\varepsilon_{\mu\alpha}^\eta\left(m_\eta^N-m_\eta(t)\right)-\sum_{\mu,\nu,\eta=1}^{d^2}h_{\mu\nu}(t)\varepsilon_{\nu\alpha}^\eta\left( m_\mu^Nm_\eta^N-m_\mu(t)m_\eta(t)\right)\\
&-\sum_{\mu,\nu,\eta=1}^{d^2}h_{\mu\nu}(t)\varepsilon_{\mu\alpha}^\eta\left( m_\eta^Nm_\nu^N-m_\eta(t)m_\nu(t)\right)+\mathcal{A}(t)[m_\alpha^N]\\
&-\sum_{\mu,\nu,\eta=1}^{d^2}\frac{b_{\mu\nu}(t)}{2}\varepsilon_{\mu\alpha}^\eta \left(m_\eta^Nm_\nu^N-m_\eta(t)m_\nu(t)\right)-\sum_{\mu,\nu,\eta=1}^{d^2}\frac{b_{\mu\nu}(t)}{2}\varepsilon_{\mu\alpha}^\eta \left(m_\nu^Nm_\eta^N-m_\nu(t)m_\eta(t)\right)\, .
\end{split}
\label{T1}
\end{equation}
First, we note that 
$$
\|\mathcal{A}(t)\left[m_\alpha^N\right]\|\le \frac{d^8c_{\rm max}\varepsilon_{\rm max}^2}{2N}\, .
$$
with $\varepsilon_{\rm max}=\sup_{\forall \mu,\nu,\eta}\left\{|\varepsilon_{\mu\nu}^\eta|\right\}$
Then, we note that the first term in Eq.~\eqref{T1} shows the difference of linear macroscopic operators and mean-field variables. All other terms involve instead the difference between quadratic objects. Since they all have the same structure, we collect all linear terms together and all quadratic terms together. In this way, we can write
\begin{equation}
\begin{split}
\mathcal{L}(t)\left[m_\alpha^N\right]-\dot{m}_\alpha(t)&=\sum_{s=1}^{d^4}q_s(t) \left(m_{\eta_s}^N-m_{\eta_s}(t)\right)
+\sum_{s=1}^{4d^6}p_s(t)\left( m_{\eta_s}^Nm_{\mu_s}^N-m_{\eta_s}(t)m_{\mu_s}(t)\right)+\mathcal{A}(t)[m_\alpha^N]\, ,
\end{split}
\label{T1-bis}
\end{equation}
where all the coefficients are such that $|q_s(t)|,|p_s(t)|\le c_{\rm max}\varepsilon_{\rm max}$. 

We can now show how to bound the different terms appearing  in Eq.~\eqref{partial}, by looking at  representative linear and nonlinear terms. For the linear terms, we consider a bound to the expectation they give rise to. This can be found by using Cauchy-Schwarz inequality and noticing that $\left\langle \Phi(t)\left[\left(m_\alpha^N-m_\alpha(t)\right)^2\right]\right\rangle \le \mathcal{E}_N(t)$. In this way, we find 
$$
\left|\left\langle \Phi(t)\left[\Big(m_{\eta_s}^N-m_{\eta_s}(t)\Big)\Big(m_\alpha^N-m_\alpha(t)\Big)\right]\right\rangle\right|\le \mathcal{E}_N(t)\, .
$$
For the nonlinear terms, we first consider that 
\begin{equation}
m_{\eta_s}^Nm_{\mu_s}^N-m_{\eta_s}(t)m_{\mu_s}(t)=\left(m_{\eta_s}^N-m_{\eta_s}(t)\right)m_{\mu_s}^N+\left(m_{\mu_s}^N-m_{\mu_s}(t)\right)m_{\eta_s}(t)\, .
\label{nonlin}
\end{equation}
Here, we observe that  $\|m_{\mu_s}^N\|\le 1$ and that, due to considerations in Lemma 1, $|m_{\eta_s}|\le \sqrt{M}$, where $M=\sum_\alpha m_\alpha^2(t)$ is a constant of motion for the mean-field equations. As such we can bound both the nonlinear terms of Eq.~\eqref{nonlin} when inserted in the first term of Eq.~\eqref{partial} as 
$$
\left|\left\langle \Phi(t)\left[\Big(m_{\eta_s}^N-m_{\eta_s}(t)\Big)X\Big(m_\alpha^N-m_\alpha(t)\Big)\right]\right\rangle\right|\le \sqrt{M}\mathcal{E}_N(t)\, ,
$$
which is valid for both $X=m_{\mu_s}$ and $X=m_{\eta_s}(t)$. The last contribution to be considered for the first term in Eq.~\eqref{partial} is given by 
$$
\left|\left\langle \Phi(t)\left[\mathcal{A}(t)\left[m_\alpha^N\right]\Big(m_\alpha^N-m_\alpha(t)\Big)\right]\right\rangle\right|\le \frac{c_{\rm max}\varepsilon_{\rm max}^2d^8}{2N}(1+\sqrt{M})\, ,
$$
where we exploited the consideration we made after Eq.~\eqref{rel2} and that $\|m_\alpha^N-m_\alpha(t)\|\le 1+\sqrt{M}$. 

All together, the above considerations show that, defining 
$$
T_1=\sum_{\alpha=1}^{d^2}\left\langle \Phi(t)\left[\Big(\mathcal{L}(t)\left[m_\alpha^N\right]-\dot{m}_\alpha(t)\Big)\Big(m_\alpha^N-m_\alpha(t)\Big)\right]\right\rangle \, ,
$$
we have 
$$
|T_1|\le c_{\rm max}\varepsilon_{\rm max} d^6\left(1+8d^2\sqrt{M}\right)\mathcal{E}_N(t)+\frac{c_{\rm max}\varepsilon_{\rm max}^2d^{10}(1+\sqrt{M})}{2N}\, .
$$
Moreover, we note that the second term $T_2$ in Eq.~\eqref{partial} is, as we already mentioned, the complex conjugate of $T_1$ so that the same bound holds also for it. Recalling also the result for $T_3$ in Eq.~\eqref{bound_3}, we overall find that 
$$
\dot{\mathcal{E}}_N(t)\le \left|\dot{\mathcal{E}}_N(t)\right|\le C_1\mathcal{E}_N(t) +\frac{C_2}{N}\, , 
$$
where 
$$
C_1= 2 c_{\rm max}\varepsilon_{\rm max} d^6\left(1+8d^2\sqrt{M}\right)\, ,\qquad C_2=c_{\rm max}\varepsilon_{\rm max}^2d^{10}(1+\sqrt{M})+4d^6 c_{\rm max}\, .
$$
Due to Gronwall Lemma, we thus have 
$$
\mathcal{E}_N(t)\le e^{tC_1}\mathcal{E}_N(0)+\frac{C_2}{C_1 N}\left(e^{tC_1}-1\right)\, 
$$
and exploiting our assumption (C2) on the initial state we find 
$$
\lim_{N\to\infty}\mathcal{E}_N(t)\le e^{t C_1}\lim_{N\to\infty}\mathcal{E}_N(0)+(e^{tC_1}-1)\lim_{N\to\infty}\frac{C_2}{C_1 N}=0\, .
$$

As a final note, we comment on the situation in which the coefficients of the dynamical generator $\omega_\mu(z),h_{\mu\nu}(z),c_{\mu\nu}(z)$ are not bounded on the whole line $t\in[0,+\infty)$. In such a case, one considers the growth of $\mathcal{E}_N(t)$ on any compact interval $t\in[0,T]$,  with arbitrary $T>0$. The analyticity  of  $\omega_\mu(z),h_{\mu\nu}(z),c_{\mu\nu}(z)$ guarantees their boundedness on  such intervals, which is enough to prove the theorem with the same argument exploited above. 
\qed

\newpage
\section*{II. Lemmata}
\begin{lemma}
Consider the generator in Eq.~\eqref{generator} for arbitrary finite $N$. Furthermore,  assume that $\omega_{\alpha}(z)$, $h_{\alpha\beta}(z)$ and $c_{\alpha\beta}(z)$ are analytic functions on a simply connected domain $D$, which contains the non-negative part of the real line, $\{z=x+iy: x\ge0, y=0\}$, in its interior. The corresponding system of mean-field differential equations in Eq.~\eqref{mf-eqs}, with initial conditions $m_\alpha(0)=m_\alpha^0$,  admits a unique (infinitely) differentiable solution for $t\in[0,\infty)$.
\end{lemma} 

{\it Proof:} The system in Eq.~\eqref{mf-eqs} is a nonlinear system of differential equations with time-dependent coefficients, which can be written as [$m=(m_1,m_2,m_3, \dots m_{d^2})^T$]
$$
\dot{m}(t)=f(t,m(t))\, .
$$
The initial time is $t=0$ and the initial conditions are given by $m_\alpha(0)=m_\alpha^0$. The vector function $f$ consists of, at most, polynomials of degree two in the $m$ variables, and is thus  infinitely differentiable with respect to $m$. The function $f$ is moreover infinitely differentiable with respect to $t$, due to our assumption on $\omega_{\alpha}(z)$, $h_{\alpha\beta}(z)$ and $c_{\alpha\beta}(z)$ being analytic in $D$ containing the non-negative real line. As such,  the vector function $f(t,m)$ is locally Lipschitz continuous in $m$, uniformly in $t$  \cite{perko13,teschl2012}. Therefore, there exists a unique local solution in the neighbourhood of $t=0$ \cite{coddington1956,teschl2012}. Due to the smoothness of the  coefficients specifying the generator, the solution $m(t)$ is an infinitely differentiable function with respect to time \cite{teschl2012}. We now want to show that such solution can be continued to the whole positive real line.  

To this end, we note that the system of differential equations features the conserved quantity 
$$
M=\sum_{\alpha=1}^{d^2}m_\alpha^2(t)=\sum_{\alpha=1}^{d^2}m_\alpha^2(0)\, ,
$$
so that one has $|m_\alpha(t)|\le \sqrt{M}$, for all times $t>0$. The vector of solutions $m(t)$ thus cannot escape any compact in $\mathbb{R}^{d^2}$ in any finite time. This implies that the solution $m(t)$ extends to the whole interval $t\in[0,+\infty)$ (see for instance Chapter $2$ of Ref.~\cite{teschl2012}). \qed \\

\begin{lemma}
Consider the generator in Eq.~\eqref{generator} for  arbitrary finite $N$. Furthermore, assume that $\omega_{\alpha}(z)$, $h_{\alpha\beta}(z)$ and $c_{\alpha\beta}(z)$ are analytic functions on a simply connected domain $D$, which contains the non-negative part of the real line, $\{z=x+iy: x\ge0, y=0\}$, in its interior. Let also $D\ni z_0=0$. Then, the map $\Phi(z)$, defined through the differential equation  $\dot{\Phi}(z)=\Phi(z)\circ \mathcal{L}(z)$, with initial condition $\Phi(z_0)={\rm id}$ (where ${\rm id}$ represents the identity map), is analytic in $D$. 
\end{lemma} 

{\it Proof:} The proof of the Lemma closely follows the standard proof for systems of ordinary equations (see, e.g., Ref.~\cite{coddington1956}, in particular Section $7$ of Chapter $3$). To this end, it is convenient to first reshape the map differential equation $\dot{\Phi}(z)=\Phi(z)\circ \mathcal{L}(z)$ into a more familiar matrix-vector differential form.\\

The Hilbert space of the considered system of $N$ $d$-level particles is spanned by a many-body basis $\{\ket{\alpha}\}_{\alpha=1}^{d^N}$. The map $\Phi(z)$ acts on operators and is thus fully specified by the quantities 
$$
\Phi_z^{(\alpha_1,\alpha_2,\alpha_3,\alpha_4)}=\bra{\alpha_3}\Phi(z)\left[\ket{\alpha_1}\!\bra{\alpha_2}\right]\ket{\alpha_4}\, ,
$$
which we can exploit to define the vector (with $\ket{\alpha_1,\alpha_2,\alpha_3,\alpha_4}=\ket{\alpha_1}\!\ket{\alpha_2}\!\ket{\alpha_3}\!\ket{\alpha_4})$
$$
 \ket{\varphi(z)}=\sum_{\alpha_1,\alpha_2,\alpha_3,\alpha_4=1}^{d^N} \Phi_z^{(\alpha_1,\alpha_2,\alpha_3,\alpha_4)}\ket{\alpha_1,\alpha_2,\alpha_3,\alpha_4}\, .
$$
The original system of differential equations thus reads 
$$
\bra{\alpha_3}\dot{\Phi}(z)\left[\ket{\alpha_1}\!\bra{\alpha_2}\right]\ket{\alpha_4}=\bra{\alpha_3}{\Phi}(z)\big[\mathcal{L}(z)\left[\ket{\alpha_1}\!\bra{\alpha_2}\right]\big]\ket{\alpha_4}\, .
$$
Introducing identities in the form $\sum_\alpha \ket{\alpha }\!\bra{\alpha}$ before and after the action of the generator, we find 
$$
\braket{\alpha_1,\alpha_2,\alpha_3,\alpha_4|\dot{\varphi}(z)}=\bra{\alpha_3}\dot{\Phi}(z)\left[\ket{\alpha_1}\!\bra{\alpha_2}\right]\ket{\alpha_4}=\sum_{\alpha_5,\alpha_6=1}^{d^N} \mathcal{L}_z^{(\alpha_1,\alpha_2,\alpha_5,\alpha_6)} \Phi_z^{(\alpha_5,\alpha_6,\alpha_3,\alpha_4)}\, .
$$
This suggests that the generator $\mathcal{L}(z)$ can be represented through the matrix $\mathbb{L}(z)$, 
$$
\mathbb{L}(z)=\sum_{\alpha_1,\alpha_2,\alpha_3,\alpha_4,\alpha_5,\alpha_6=1}^{d^N} \mathcal{L}_z^{(\alpha_1,\alpha_2,\alpha_5,\alpha_6)}\ket{\alpha_1,\alpha_2,\alpha_3,\alpha_4}\bra{\alpha_5,\alpha_6,\alpha_3,\alpha_4}\, ,
$$
and that the differential equation for the map $\Phi(z)$ can be converted into the more familiar looking form 
\begin{equation}
\ket{\dot{\varphi}(z)}=\mathbb{L}(z)\ket{\varphi(z)}\, .
\label{IVP}
\end{equation}
Since the map $\Phi({z=0})$ is the identity map, we have $\Phi_{0}^{\alpha_1,\alpha_2,\alpha_1,\alpha_2}=1$, $\forall \alpha_1,\alpha_2$,  with all other entries being zero. We can now exploit standard arguments for ordinary differential equations \cite{coddington1956}. \\

Due to our assumption on the dynamical generator $\mathcal{L}(z)$, the matrix $\mathbb{L}(z)$ (which is finite dimensional) is analytic within a simply connected domain $D$. We consider a point $z_1$ inside the domain and a smooth arc $\Gamma$, from $z_0$ to $z_1$, of overall length $\ell$. Since $\mathbb{L}(z)$ is analytic in $D$, it is possible to find a constant $C$, such that $\|\mathbb{L}(z)\|<C$ for $z\in \Gamma$. We construct an approximation to the solution of the differential equation, essentially through truncated Dyson series, as
$$
\ket{\varphi^n(z)}=\sum_{m=0}^n \int_0^z {\rm d}s_n\int_0^{s_n} {\rm d}s_{n-1}\dots \int_0^{s_2}{\rm d}s_1 \mathbb{L}({s_n})\mathbb{L}({s_{n-1}})\dots \mathbb{L}({s_1})\ket{\varphi(0)}\, ,
$$
where integration is taken along the arc $\Gamma$. These functions are analytic and satisfy the iterative equation
\begin{equation}
\ket{\varphi^n(z)}=\ket{\varphi(0)}+\int_0^z {\rm d}s \, \mathbb{L}(s) \ket{\varphi^{n-1}(s)}\, . 
\label{iterative_relation}
\end{equation}
The approximation functions are  bounded by 
$$
\left\|\ket{\varphi^n(z)}\right\|\le \|\ket{\varphi(0)}\|\sum_{m=0}^n \frac{C^n \ell^n}{n!} \, .
$$
The above argument is actually valid for any point $z\in D$ which can be reached from $z_0$ and is thus valid for any closed region $R$ in $D$. The functions $\ket{\varphi^n(z)}$ are therefore analytic in $R$ and they uniformly converge to the analytic, in $R$, function $\ket{\varphi(z)}=\lim_{n\to\infty}\ket{\varphi^n(z)}$. Looking at the relation in Eq.~\eqref{iterative_relation}, it can be seen that $\ket{\varphi(z)}$ is the (unique) solution of the differential equation in Eq.~\eqref{IVP}. Given that the functions in the vector $\ket{\varphi(z)}$ are analytic in $D$, the map $\Phi(z)$ is also analytic in such domain. 
\qed 

\newpage
\section*{III. Dissipative Floquet Ising model}
In this Section we provide details on the dissipative Floquet Ising model that we consider in the main text. The dynamical generator, in diagonal form, for such a system is given by the map
\begin{equation}
    \mathcal{L}(t)[X]=i[H(t), X]+\gamma \Big(2V_+X V_--V_+V_-X-XV_+V_- \Big)\, .
\end{equation}
Here, the Hamiltonian $H(t)$ is the one given in Eq.~\eqref{Ising_Hamiltonian} and the operators appearing in the dissipative part of the generator are $V_-=(V_1-iV_2)/(2\sqrt{N})$ and $V_+=V_-^\dagger$. Due to the collective character of the dynamics, the total angular momentum $V^2=V_1^2+V_2^2+V_3^2$ is dynamically conserved. We focus on the fully symmetric subspace in which $V^2$ is maximal.

For the above system, the mean-field equations read 
\begin{equation}
    \begin{split}
        \dot{m}_1(t)&=-2\left[\Delta_0+\Delta_1 \sin \chi t\right]m_2(t)+\gamma m_1(t) m_3(t)\, ,\\
        \dot{m}_2(t)&=4g m_1(t) m_3(t)+2\left[\Delta_0+\Delta_1 \sin \chi t\right]m_1(t)+\gamma m_2(t) m_3(t) \, , \\
        \dot{m}_3(t)&=-4g m_1(t) m_2(t) -\gamma [m_1^2(t)+m_2^2(t)]\, .
    \end{split}
\end{equation}
Conservation of the total angular momentum to its maximal value  requires $m_1^2(t)+m_2^2(t)+m_3^2(t)=1$ at the level of the mean-field equations. 
We solve the above equations numerically with initial conditions associated with the state $\ket{\psi_0}=\bigotimes_{k=1}^N\ket{\theta,\phi}$. That is, $m_1(0)=\sin \theta \cos \phi$, $m_2(0)=\sin \theta \sin \phi$, and $m_3(0)=\cos \theta$. 
With regard to  the finite-$N$ numerical simulations, we generate the initial state as 
$$
\rho(0)=\ket{\psi(0)}\!\bra{\psi(0)}\, , \qquad \mbox {with } \qquad \ket{\psi(0)}=e^{-\frac{i}{2}V_3\phi}e^{-\frac{i}{2} V_2 \theta} \ket{N}\, ,
$$
and  $V_3\ket{N}=N\ket{N}$. The quantum state is evolved by considering the dual propagator $\Phi^*(t)$, such that $\rho(t)=\Phi^*(t)[\rho(0)]$. The latter obeys the differential equation $\dot{\Phi}^*(t)=\mathcal{L}^*(t)\circ {\Phi}^*(t)$, with dual Lindblad generator (in the Schr\"odinger picture)
$$
\mathcal{L}^*(t)[\rho]=-i[H(t),\rho]+\gamma \Big(2V_-\rho V_+-V_+V_-\rho -\rho V_+V_-\Big)\, .
$$

To obtain the gap of the Floquet generator $\Phi(\tau)$, we consider the map $\Phi^*(\tau)$ ---which possesses the same spectrum of $\Phi(\tau)$---  and approximate it, using a Trotter expansion, as  (with $M=5000$)
$$
\Phi^*(\tau)\approx \prod_{k=1}^{M} \left(e^{\frac{\tau}{M}\mathcal{L}_1^*\left(\frac{k\tau}{M}\right)}\circ e^{\frac{\tau}{M}\mathcal{L}_0^*}\right)\, .
$$
Here, the product means composition of maps and it is ordered with the smallest $k$ on the right and the largest ones on the left. Moreover, we have defined the time-dependent map
$$
\mathcal{L}_1^*(t)[\rho]=-i[\Delta_1 \sin (\chi t) V_3,\rho]\, , 
$$
from which we also have the time-independent term  $\mathcal{L}_0^*=\mathcal{L}^*(t)-\mathcal{L}_1^*(t)$.

\end{document}